# Tailoring electronic properties of multilayer phosphorene by siliconization

Oleksandr I. Malyi[1*], Kostiantyn V. Sopiha[2*], Ihor Radchenko[2], Ping Wu[2] and Clas Persson[1]

1 – Centre for Materials Science and Nanotechnology, Department of Physics, University of Oslo, P. O. Box 1048 Blindern, NO-0316 Oslo, Norway

2 – Engineering Product Development, Singapore University of Technology and Design, 8 Somapah Road, 487372 Singapore, Singapore

email: oleksandrmalyi@gmail.com (O.I.M)

*these authors contributed equally

**Abstract**

Controlling a thickness dependence of electronic properties for two-dimensional (2d) materials is among primary goals for their large-scale applications. Herein, employing a first-principles computational approach, we predict that Si interaction with multilayer phosphorene (2d-P) can result in the formation of highly stable 2d-SiP and 2d-SiP$_2$ compounds with a weak interlayer interaction. Our analysis demonstrates that these systems are semiconductors with band gap energies that can be governed by varying the thicknesses and stacking arrangements. Specifically, siliconization of phosphorene allows to design 2d-SiP$_x$ materials with significantly weaker thickness dependence of electronic properties than that in 2d-P and to develop ways for their tailoring. We also reveal the spatial dependence of electronic properties for 2d-SiP$_x$ highlighting the difference in effective band gaps for different layers. Particularly, our results show that central layers in the multilayer 2d systems determine their overall electronic properties, while the role of the outermost layers is noticeably smaller.

**Introduction**

Since a single layer of black phosphorus named phosphorene had been mechanically exfoliated from its bulk counterpart,[1, 2] it has received considerable attention due to its unique properties. It is known that phosphorene is a direct or near direct band gap semiconductor[3-6] which already demonstrated a great potential for metal-ion batteries,[7, 8] transistors,[1, 9, 10] chemical sensors,[11] etc. Extensive efforts were also made to understand and control the electron transport in phosphorene by introducing mechanical stress and various defects/chemical doping.[12-20] Recently, a strong dependence of electronic properties for 2d-P on its thickness (i.e., number of phosphorene layers) was predicted using hybrid functional first-principles calculations.[3, 4] It was shown that the energy gap of 1.52 eV for a single-layer phosphorene drops to 1.02 eV for the bilayer, and further decreases rapidly with thickness until the bulk value of 0.39 eV is reached.[4] Later, the computed dependence was confirmed experimentally using the evolution of polarization-resolved reflection spectra for phosphorene-based systems.[5] On the one hand, this relationship opens a possibility to synthesize a variety of 2d-P materials with tunable electronic properties. On the other hand, it may limit the applications of 2d-P in cases when a thick active material with a wide energy gap is needed.

Similar to other 2d systems, controlling 2d-P thickness during synthesis is difficult at the nanoscale. For example, standard methods of mechanical and liquid exfoliations can only produce a random mixture of black phosphorus flakes with different thicknesses, where the single-layer flakes are notably rare.[21-23] Although recent advances in mechanical exfoliation technique allow increasing the yield of few-layer black phosphorus,[24, 25] accurate control of its thickness remains challenging. Several research groups have also reported a direct high-pressure synthesis of black phosphorus films from the CVD-grown red phosphorus.[26, 27] However, the reported film thicknesses of 40–200 nm suggest the formation of polycrystalline black phosphorus rather than a single-layer phosphorene. All these complications in the synthesis make it difficult to predict the performance of 2d-P-based devices as the thickness of 2d-P and, consequently, its electronic properties may vary significantly among the samples. Hence, it is essential to establish a simple but effective mechanism to reduce the thickness dependence of energy gaps in 2d-P. Currently, several approaches (e.g., intercalating atoms,[28] change of stacking,[29] and applying electric field[30, 31]) for tuning electronic properties of 2d materials exist. In this work, we propose an alternative strategy utilizing phosphorene reactivity to form highly stable 2d-SiP$_x$ materials with weaker thickness dependence of electronic properties than that in 2d-P.

**Methods**

Density functional theory (DFT) calculations are carried out using the Vienna Ab Initio Simulation Package (VASP).[32-34] Since the investigation of layered materials requires an accurate description of van der Waals (vdW) interactions, all presented calculations are performed using several different treatments of exchange-correlation (XC) functional that include different models of the vdW interactions. Specifically, we use the Perdew-Burke-Ernzerhof (PBE, which does not account vdW interaction),[35] DFT-D2,[36] and vdW-DF (optPBE-vdW, optB88-vdW, and optB86b-vdW)[37, 38] functionals. Herein, the application of the multiple XC functionals is vital to reveal the sensitivity of results to the used setup as some limitations of existing vdW approaches in the description of 2d materials have recently been pointed out.[39] To predict the electronic properties accurately, we also perform calculations using the revised Heyd-Scuseria-Ernzerhof (HSE) screened hybrid functional.[40] Here since HSE does not capture the vdW interaction itself, the atomic positions and lattice vectors are kept the same as those in the lowest energy structures found for each XC functional. The cutoff energies for plane-wave basis are set to 300 and 400 eV for the Born-Oppenheimer molecular dynamics (BOMD) simulations and for all other DFT calculations, respectively. Γ-centered

Monkhorst-Pack k-grids[41] of various sizes are used in the Brillouin-zone integrations (see Table S1). Atomic relaxations are carried out until the internal forces are smaller than 0.01 eV/Å. To evaluate system stability, the interlayer interaction energy per atom ($E_{inter}$) is calculated as $E_{inter} = (E_s - nE_l)/k$, where $E_s$ is the energy of multilayer system containing $k$ atoms arranged in $n$ 2d layers, and $E_l$ is the energy of the isolated 2d layer. We also compute Bader charges with respect to number of valence electrons using the approach developed by Henkelman et al.[42-44] To understand electronic properties of the studied 2d materials better, layer-resolved local density of states (LDOS) and integrated layer-resolved LDOS are calculated. The LDOS is estimated by adding atom-projected density of states for all atoms in each layer. Although this method neglects the density of states outside the Wigner-Seitz cells, the LDOS serves as a simple descriptor for the contribution of each layer to the overall electronic properties. Since the accurate LDOS calculation requires dense k-grid, these quantities are presented for optB88-vdW functional. We also show the HSE results with sparser k-point samplings in Supplementary Information. The obtained results are analyzed using Vesta[45] and pymatgen.[46]

**Results and discussion**

Electronic properties of 2d-P have a strong thickness dependence. As an illustration, for the 2d-P structure relaxed using optB88-vdW functional, the HSE energy gap varies from 1.52 eV for a single-layer phosphorene to 0.38 eV for black phosphorus as shown in Tables S2-S3, which is in good agreement with previous studies.[3-5] However, at the first-principles level, the thickness dependence is sensitive to the XC functional used for the structure relaxation as shown in Fig. 1a. In particular, since PBE neglects non-local correlations, the interlayer interaction is strongly underestimated. This results in a significantly larger separation between the phosphorene layers compared to those for other used functionals. For bilayer phosphorene, the interlayer distance and interaction energy for optB86b-vdW are 3.09 Å and -59 meV/atom, respectively, while the corresponding PBE values are 3.50 Å and -5 meV/atom (see Figs. 1b-d). By studying structural and electronic properties of the bilayer phosphorene, we find that the variation in the thickness dependence is in large part due to a change of the interlayer distance. Here the energy gap monotonically follows the interlayer distance reaching the single-layer phosphorene limit at the large separation (see Fig. 1d). Hence, this correlation sheds light on the diversity of gap energies reported in previous works (see Ref. [47] and references therein). However, the interlayer distance is not the only parameter determining the thickness dependence as the variation of HSE energy gap is also found for a single-layer phosphorene from various XC functionals as shown in Table S3. For example, for the optB86b-vdW and PBE structures, the HSE energy gaps are 1.46 and 1.59 eV, respectively. This divergence is mainly due to in-plane relaxation. Although the average P-P bond length is 2.24±0.01 Å for all XC functionals, the difference in lattice constants along the armchair direction reaches 3%, while the corresponding changes in other lattice directions are only about 1%. Considering typical differences in the structural parameters, the variation in the energy gap can be understood from the superposition of different lattice strains.[14, 18] These results also highlight the need to develop next-generation of XC functionals for a better description of not only interlayer distances but also interaction energies and proper description of in-plane relaxations.

Since the interaction between the phosphorene layers has a major contribution to the thickness dependence, we hypothesize that its governing can be used to tailor the electronic properties. Because of high phosphorene reactivity, we believe that the dependence can be governed via controlled formation of stable 2d P-based materials. Based on screening of binary P-based systems available in Material Project database,[48] we select $SiP_x$ compounds due to their stability and possibility to form energetically stable 2d materials[49, 50] retaining the waved feature of experimentally synthesized bulk SiP$_x$ systems. It should be noted that alternative structures of single-layer 2d-SiP$_x$ were proposed in recent studies,[50-53] but these structures are not considered herein due to their instability or/and limited information of their synthesis even in bulk SiP$_x$ forms.

Precipitation of SiP microcrystals in heavily P-doped Si matrix was originally reported by Schmidt and Stickler[54] as well as by ÓKeeffe, et al.[55] Shortly thereafter, Wadsten was the first to identify the space group of orthorhombic SiP phase as *Pnna*,[56] which was changed to *Cmc*2$_1$ in his later work.[57] Since then, successful synthesis of SiP using various growth methods has been reported by multiple research groups.[58-63] In most cases, the crystals were grown in the orthorhombic phase,[58-60] however, some studies reported a concurrent growth of both the orthorhombic and monoclinic *C*2*/m* phases,[61, 62] indicating their close energetics. Indeed, the difference in optB86b-vdW formation heats of these two phases is smaller than 0.2 meV/atom. This is expected as the phases consist of nearly-identical SiP layers combined by long-range vdW interaction with distinct stacking arrangements as demonstrated in Figs. 2a,b. Here the stability of the single layer is determined by covalent Si-Si and polar covalent P-Si bonds. The bonding nature is well illustrated by average Bader charges of -1.14e and 1.14e on P and Si (the data are given for single-layer 2d-SiP computed using PBE functional), respectively. The Si-P interaction is sufficient to stabilize the single-layer 2d-SiP with respect to bulk Si and black P by the PBE formation heat of -0.17 eV/atom. For all considered XC functionals except optB88-vdW, 2d-SiP(*Cmc*2$_1$) structures are slightly more stable than 2d-SiP(*C*2*/m*) (for simplicity, the 2d materials are labeled by the space group names of the bulk phases, see Figs. 2c,d). The energy difference between the 2d-SiP(*Cmc*2$_1$) and 2d-SiP(*C*2*/m*) is smaller than 1 meV/atom. It indicates that both stacking arrangements might be realized at the nanoscale. 2d-SiP has consistently weaker interlayer interaction than 2d-P as summarized in Tables S4-S5. These results suggest that the isolation of single-layer 2d-SiP is highly likely.

Single-layer 2d-SiP is a semiconductor with HSE energy gap of 2.66±0.01 eV. In contrast to phosphorene, due to the nearly-identical structure geometries, the computed gap energy is not sensitive to the choice of XC functional for the relaxation. It should be noted that since SiP is an indirect band gap material (the same holds for $SiP_2$ discussed below), there is a minor overestimation of the energy gap reaching 0.1 eV due to the finiteness of the used k-points grids. Because of this and high computational cost of HSE calculations, herein, the HSE values computed with the sparser k-point samplings are corrected by the difference of corresponding vdW-DF (PBE or DFT-D2) energy gaps for the same and denser k-point grids (see Table S1). When the number of SiP layers increases, the system description becomes even more complex. First, there is an apparent variation in the interlayer distance for the used XC functionals affecting the thickness dependence of electronic properties. For instance, for bilayer 2d-SiP($Cmc2_1$), the HSE energy gap correlates monotonically with the interlayer distance as demonstrated in Fig. S1. Second, there is an additional degree of freedom to vary the electronic properties by stacking arrangements (i.e., the formation of 2d-SiP($Cmc2_1$) and 2d-SiP($C2/m$) structures). In particular, for optB88-vdW relaxed structures, the difference in HSE energy gaps of bilayer 2d-SiP($Cmc2_1$) and 2d-SiP($C2/m$) systems exceeds 0.2 eV. Evidently, it is not caused by a change in the structural parameters of the SiP layers in the bilayer structures, as the same energy gaps of isolated non-relaxed SiP layers for both systems are found. In fact, the variation can be understood from the analysis of the shortest interatomic distances between the atoms belonging to neighboring SiP layers (P-P distance for these structures). For example, the DFT-D2 distances are 3.53 and 3.58 Å for 2d-SiP($C2/m$) and 2d-SiP($Cmc2_1$), respectively. These values correlate with the HSE energy gaps of 2.01 and 2.19 eV for the corresponding structures. The same tendency is also found for the structures from the other XC functionals. For both 2d-SiP($C2/m$) and 2d-SiP($Cmc2_1$), the increase in the number of SiP layers reduces the energy gaps (see Figs. 2e,f). The change in band gap energy is comparable to that for 2d-

P. For instance, for optB88-vdW relaxed structures of 2d-P, 2d-SiP(*Cmc*2$_1$), and 2d-SiP(*C2/m*), the difference in energy gaps between the single- and 5-layer systems are 0.91, 0.75, and 1.00 eV, respectively. Regardless of the XC functional used in the structure relaxation, this change is the smallest for 2d-SiP(*Cmc*2$_1$). At the same time, for 2d-P and 2d-SiP(*C2/m*), it is not possible to identify which material has the smaller change in energy gap as different trends are found for the considered XC functionals (see Figs. 2e,f). Anyway, these results reveal a great potential of phosphorene siliconization to form a new generation of 2d materials and to control their properties by tuning their stacking arrangements and thicknesses.

From a thermodynamic perspective, the Si-P interaction can also stabilize layered P-rich SiP$_2$ compounds. However, precipitation of SiP$_2$ in P-doped Si matrix has not been observed experimentally due to the competitive growth of Si-rich SiP phases. Up to date, a few successful syntheses of two distinct layered SiP$_2$ structures were reported.[64-66] First, Wadsten synthesized needle-like crystals of SiP$_2$ from a stoichiometric mixture of the pure elements and identified their crystal structure as orthorhombic *Pbam* using X-ray diffraction (XRD) techniques.[64] Later, SpringThorpe grown SiP$_2$ crystals with a similar morphology using solution synthesis in molten Sn containing a considerable amount of Mg.[65] An akin approach but using Gd as a mineralizer has been recently utilized to synthesize another orthorhombic SiP$_2$ phase with *Pnma* crystal symmetry.[66] The possibility to grow two different SiP$_2$ phases by slight variation of the synthesis conditions indicates their close energetics. It is not surprising, similar to SiP, both orthorhombic SiP$_2$ phases consist of nearly-identical vdW-bonded 2d sheets with distinct stacking arrangements as illustrated in Figs. 3a,b. Our first-principles results suggest that the *Pnma* phase is thermodynamically more stable than the *Pbam* one. The same tendency is also observed for 2d-SiP$_2$ isolated from the bulk systems. Although the difference in the formation energies for two stacking arrangements is larger than that in 2d-SiP, it is still in the order of meV/atom. Because of this, and since the interlayer interaction energy is comparable to that for 2d-P and 2d-SiP (see Figs. 3c,d), both systems might be possible to synthesize.

The isolated single 2d-SiP$_2$ layer is a semiconductor with polar covalent Si-P bonds (for PBE relaxed structure, the average Bader charges of Si and P are 1.52e and -0.76e, respectively) and the HSE energy gap varying from 2.20 to 2.33 eV depending on the XC functional used in the relaxations. Akin to phosphorene, the variation in the energy gap of single-layer 2d-SiP$_2$ is associated with the in-plain layer geometry. For instance, the difference in lattice constants along the wave direction reaches 3.5%, while along other lattice directions it is about 1%. Here the HSE energy gap follows monotonically the lattice constant along the wave direction for structures computed using various XC functionals (see Tables S6-S7).

The increase in 2d-SiP$_2$ thickness leads to a noticeable change in the electronic properties. The largest difference in HSE energy gap of single- and 5-layer 2d-SiP$_2$ systems reaches 0.43 eV (the data are given for optB86b-vdW relaxed structures). As summarized in Tables S6-S7, 2d-SiP$_2$(*Pnma*) structures have slightly larger HSE energy gaps than those of 2d-SiP$_2$(*Pbam*) systems for all considered thicknesses. Similar to 2d-SiP, the effect of stacking arrangement on electronic properties can be understood from the analysis of the shortest interatomic distances for atoms belonging to the neighboring 2d-SiP$_2$ layers. The difference in electronic properties of 2d-SiP$_2$(*Pnma*) and 2d-SiP$_2$(*Pbam*) structures increases with the thickness as shown in Figs. 3e,f. In particular, for DFT-D2 relaxed structures, the difference in energy gaps for the bilayer systems is 0.05 eV, while for the 5-layer structures it is 0.11 eV.

To better understand the electronic properties of the considered 2d-SiP$_x$ and their link with the materials symmetry, we analyze the layer-resolved LDOS and integrated layer-resolved LDOS (see Figs. 4a-d, S2-S3, and Table S8). Depending on even or odd number of layers, 2d-P has *Pbma* or *Pman* layer group symmetry, respectively. According to both symmetry and integrated LDOS analyses, there are three nonequivalent phosphorene layers in the 5-layer system. Here the central layer determines both valence and conduction band edges, while the corresponding role of the outermost layers is the smallest. These results are also reflected in the integrated LDOS shown in Fig. 4b. The same tendency is found for the outermost and central layers for other considered thicknesses. In general, the computed trends imply that the central layer determines the overall electronic properties and the thickness dependence of energy gap is mainly caused by the interlayer interaction resulting in broadening of valence and conduction bands. This is well illustrated in Fig. S2 by comparison the layer-resolved LDOS of outermost and central layers for different thicknesses. Similar tendencies are also observed for 2d-SiP(*Cmc*$2_1$) and 2d-SiP$_2$(*Pnma*) as shown in Figs. 4c,d and Fig. S3. However, unlike other structures, each layer of 2d-SiP(*Cmc*$2_1$) has distinct integrated LDOS (see Fig. 4c). Indeed, multilayer 2d-SiP(*Cmc*$2_1$) has *Cm*11 layer group symmetry and is not symmetrical in respect to the central layer. Because of this, each layer contributes differently to the overall electronic properties. Despite this, for 5-layer 2d-SiP(*Cmc*$2_1$), both the outermost layers have significantly smaller contributions near the band edges than the central layer.

Among the considered systems, 2d-SiP$_2$(*Pnma*) has the weakest variation in its integrated LDOS (see Fig. 4d). From the symmetry analysis, 2d-SiP$_2$(*Pnma*) has *P*$2_1$/*m*11 or *Pm*$2_1$*b* layer group symmetry depending on the even or odd number of layers in the system, respectively. Because of this, similar to 2d-P, three unique layers are found in 5-layer systems. Here the contributions of the central and second outer layers to the electronic properties are comparable, and thus only the outermost layers have the distinct properties. This result reflects the weak dependence of electronic properties for 2d-SiP$_2$(*Pnma*), where the difference between the energy gaps of the single- and 5-layer systems is only 0.22 eV (the value is given for optB88-vdW relaxed structures). The spatial dependence of electronic properties suggests a complex electron transport mechanism in the material. In particular, the smaller integrated LDOS of the outermost layers at both valence and conduction bands implicitly indicates their limited contributions to the electron conductivity. The comparison of the integrated layer-resolved LDOS for different layers can also be used to estimate the spatial dependence of free carrier concentration versus quasi Fermi level in degenerate materials. For 5-layer systems, at 0.1 eV below the valence band maximum (VBM), the relative carrier concentrations computed as the ratio of the integrated LDOS for the outermost and central layers are 0.4 and 0.8 for 2d-P and 2d-SiP$_2$(*Pnma*), respectively. Simultaneously due to the asymmetry of 2d-SiP(*Cmc*2$_1$), there are two different outermost layers with the relative carrier concentrations of 0.2 and 0.1. The similar tendency is also observed for quasi Fermi level above the conduction band minimum (CBM). To illustrate the difference in electron/hole behaviors for each layer, we determine effective band gap energy for each layer calculated from the integrated layer-resolved LDOS by neglecting 0.5 and 3 me/atom. Although it is sensitive to the threshold, the noticeable difference in electronic properties for the outermost and central layers is found for all considered systems (see Figs. 4e-g). Here the effective energy gap of the central layer decreases with the number of layers monotonically. In contrast, the dependence is much weaker for the outermost layers. Indeed, for the 3 me/atom threshold, there is no change in the effective energy gap between the bilayer and to 5-layer systems for the outermost layers. These observations indicate that phosphorene siliconization can be used not only to modify the thickness dependence of band gap energy but also to tune the role of the individual layers in the electronic properties.

It is well-known that some 2d materials are unstable with respect to structural perturbations and tend to reconstruct into more stable forms.[13] In order to ensure the materials stability, we also perform BOMD simulations for both single-layer 2d-SiP and 2d-SiP$_2$ at temperatures of 300 and 600 K using the PBE functional. For each BOMD simulations, we extract one system for further stability analysis every 5 ps. More details on such approach are given in our previous works.[13, 67] As it can be seen in Fig. 5a, the potential energy remains roughly constant during the simulations, and thus, no formation of new structures is observed. This is also consistent with Fig. 5b showing PBE formation heats for structures taken from the BOMD simulations suggesting high stability of the isolated 2d-SiP$_x$ layers. Because of this and negative formation heats for 2d-SiP and 2d-SiP$_2$, it is possible to utilize phosphorene siliconization or exfoliation technique (see Figs. 5c,d) to synthesize 2d materials with significantly weaker thickness dependences of electronic properties as illustrated in Figs. 5e,f. These results demonstrate a great potential for application of 2d-SiP$_x$ in various semiconductor technologies, including optoelectronics, transistors, sensors, etc.

**Conclusions**

Using first-principles calculations, we performed a detailed analysis of the stability and electronic properties of 2d-P, 2d-SiP, and 2d-SiP$_2$ considering the thickness, stacking arrangements, and different ways to describe the vdW interaction. We demonstrated that siliconization of 2d-P can result in the formation of highly stable 2d-SiP$_x$ materials having much weaker thickness dependences of electronic properties. At first-principles level, however, the electronic properties and interlayer interaction are sensitive to XC functional used for the relaxation. For all considered approaches, the thickness dependences of energy gaps for 2d-SiP and 2d-SiP$_2$ systems are noticeably weaker than that for 2d-P. In particular, for optB88-vdW relaxed structures of 2d-SiP(*Cmc*2$_1$) and 2d-SiP$_2$(*Pnma*), the difference in HSE band gap energy between the single- and 5-layer systems are 0.75 and 0.22 eV, respectively, which are significantly smaller compared to that for 2d-P (0.91 eV). Apart from that, we demonstrate the spatial dependence of electronic properties revealing that the outermost and central layers have clear difference in the effective band gaps. For the outermost layers, the effective energy gaps estimated from the integrated layer-resolved LDOS at 3 me/atom threshold do not change drastically from the bilayer to 5-layer systems. In contrast, the effective energy gaps of the central layers depend on thickness monotonically. Based on the obtained results, we predict that siliconization of phosphorene can be utilized to design novel multilayer 2d materials with not only modified thickness dependence of band gap energy but also the tunable role of each individual layer in the electronic properties.

**Acknowledgments**

This work is financially supported by the Research Council of Norway (ToppForsk project: 251131). We acknowledge PRACE for awarding access to resource MareNostrum based in Spain at BSC-CNS and HPC resources from Abel in Norway at UiO, operated by USIT, and allocation provided through NOTUR.

**Electronic Supplementary Information (ESI) available:** details on the modeled systems, summarized results, additional figures.

**Figures and Captions**

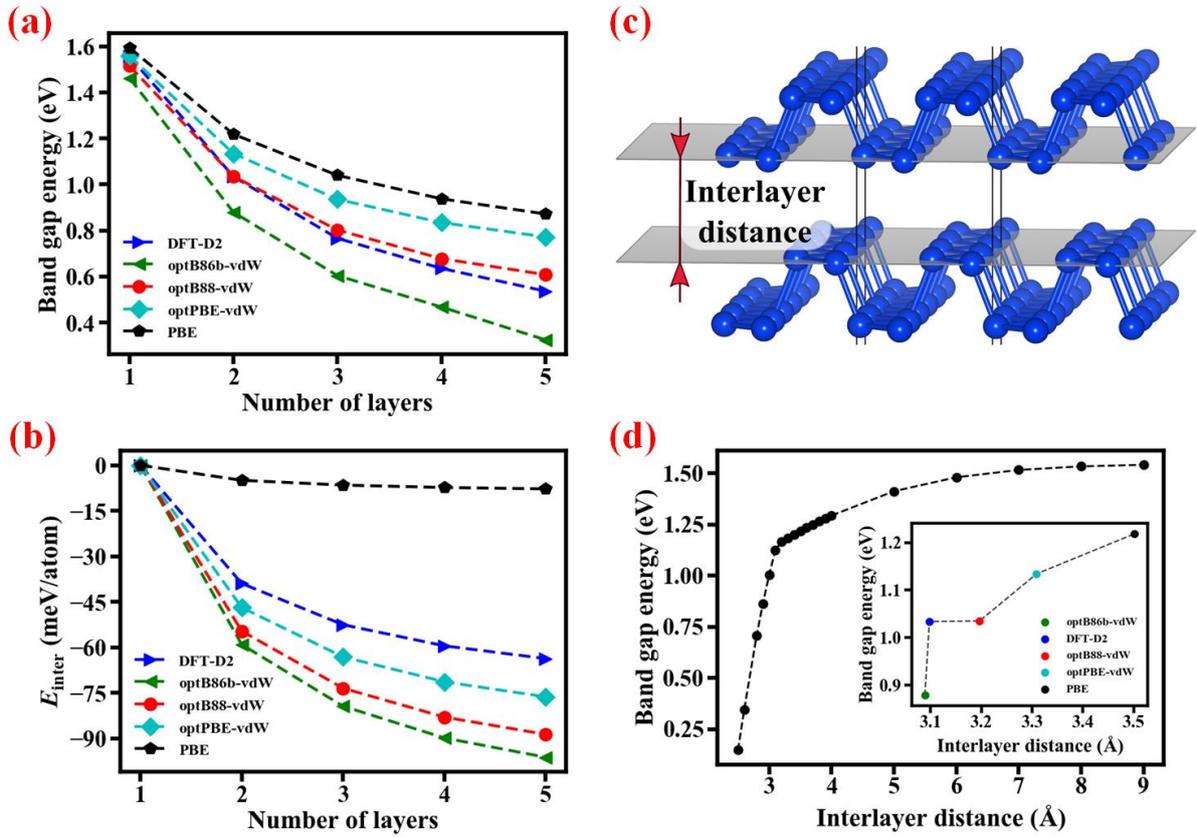

**Figure 1** (a) Thickness dependence of HSE energy gap for multilayer phosphorene relaxed with different XC functionals. (b) Interlayer interaction energy ($E_{\text{inter}}$) as a function of thickness. (c) Schematic illustration of interlayer distance and (d) HSE energy gap as a function of the interlayer distance in bilayer phosphorene; the calculations are performed for the PBE-relaxed structures. The inset in (d) shows the HSE energy gaps for the structures obtained using different XC functionals and the corresponding interlayer distances.

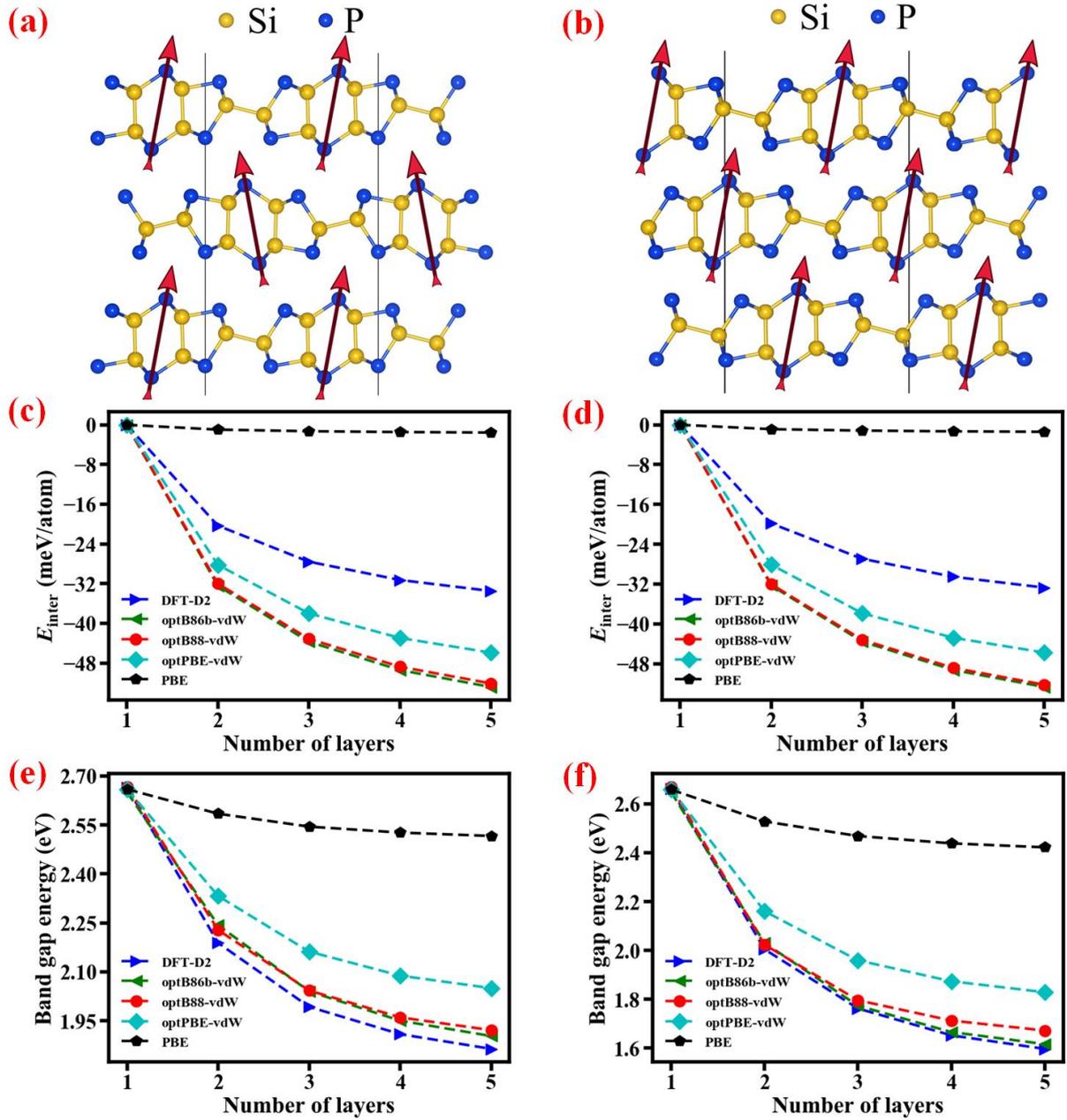

**Figure 2** Schematic illustration of two different stacking arrangements for 2d-SiP isolated from bulk (a) SiP($Cmc2_1$) and (b) SiP($C2/m$). Interlayer interaction energies ($E_{inter}$) for (c) 2d-SiP($Cmc2_1$) and (d) 2d-SiP($C2/m$) systems as functions of their thicknesses. Thickness dependences of HSE energy gaps for (e) 2d-SiP($Cmc2_1$) and (f) 2d-SiP($C2/m$) structures relaxed with different XC functionals.
17

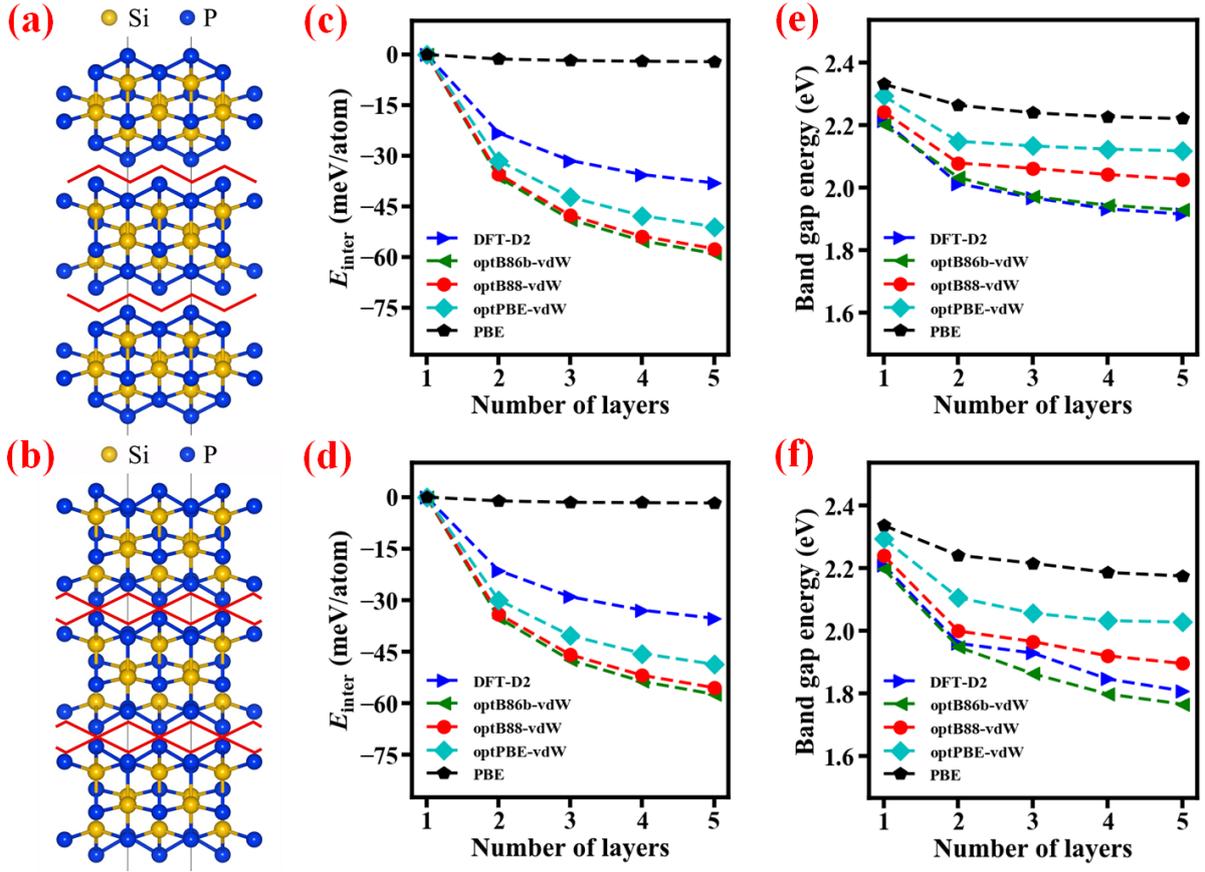

**Figure 3** Schematic illustration of two different stacking arrangements for 2d-SiP$_2$ isolated from bulk (a) SiP$_2$(*Pnma*) and (b) SiP$_2$(*Pbam*). Interlayer interaction energies ($E_{inter}$) for (c) 2d-SiP$_2$(*Pnma*) and (d) 2d-SiP$_2$(*Pbam*) systems as functions of their thicknesses. Thickness dependences of HSE energy gaps for (e) 2d-SiP$_2$(*Pnma*) and (f) 2d-SiP$_2$(*Pbam*) structures relaxed with different XC functionals.

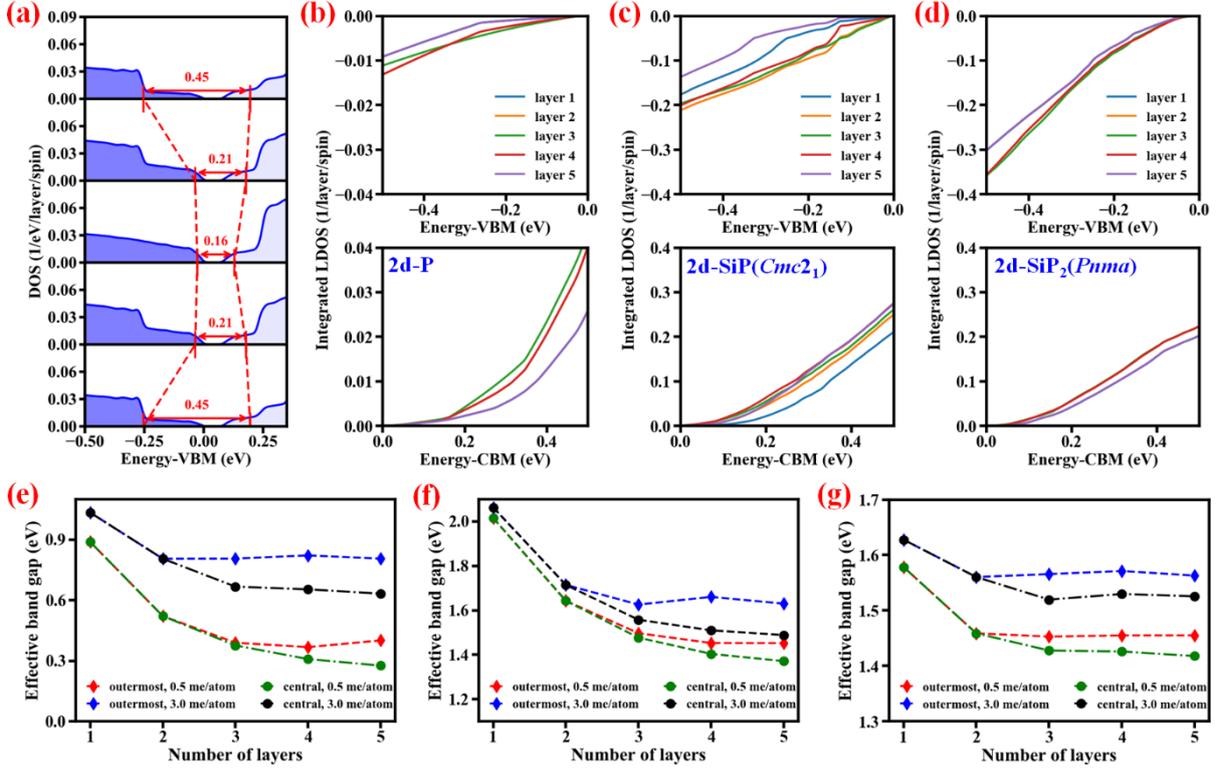

**Figure 4** (a) Layer-resolved local density of states (LDOS) for 2d-P. The difference in the LDOS is highlighted by determining the effective energy gap for each layer by neglecting densities below 0.01 1/eV/layer/spin. Integrated layer-resolved LDOS for (b) 2d-P, (c) 2d-SiP(*Cmc*2$_1$), and (d) 2d-SiP$_2$(*Pnma*). The integrated LDOS is calculated by integration of the LDOS with respect to that for VBM (or CBM). The layer index increases from bottom to top (see Fig. S4). Thickness dependence of the effective band gap energy computed from the integrated layer-resolved LDOS with 0.5 and 3 me/atom thresholds for (e) 2d-P, (f) 2d-SiP(*Cmc*2$_1$), and (g) 2d-SiP$_2$(*Pnma*). For 2d-SiP(*Cmc*2$_1$), the results are given for averaged integrated layer-resolved LDOS of the outermost/central layers. All results are computed using optB88-vdW functional.

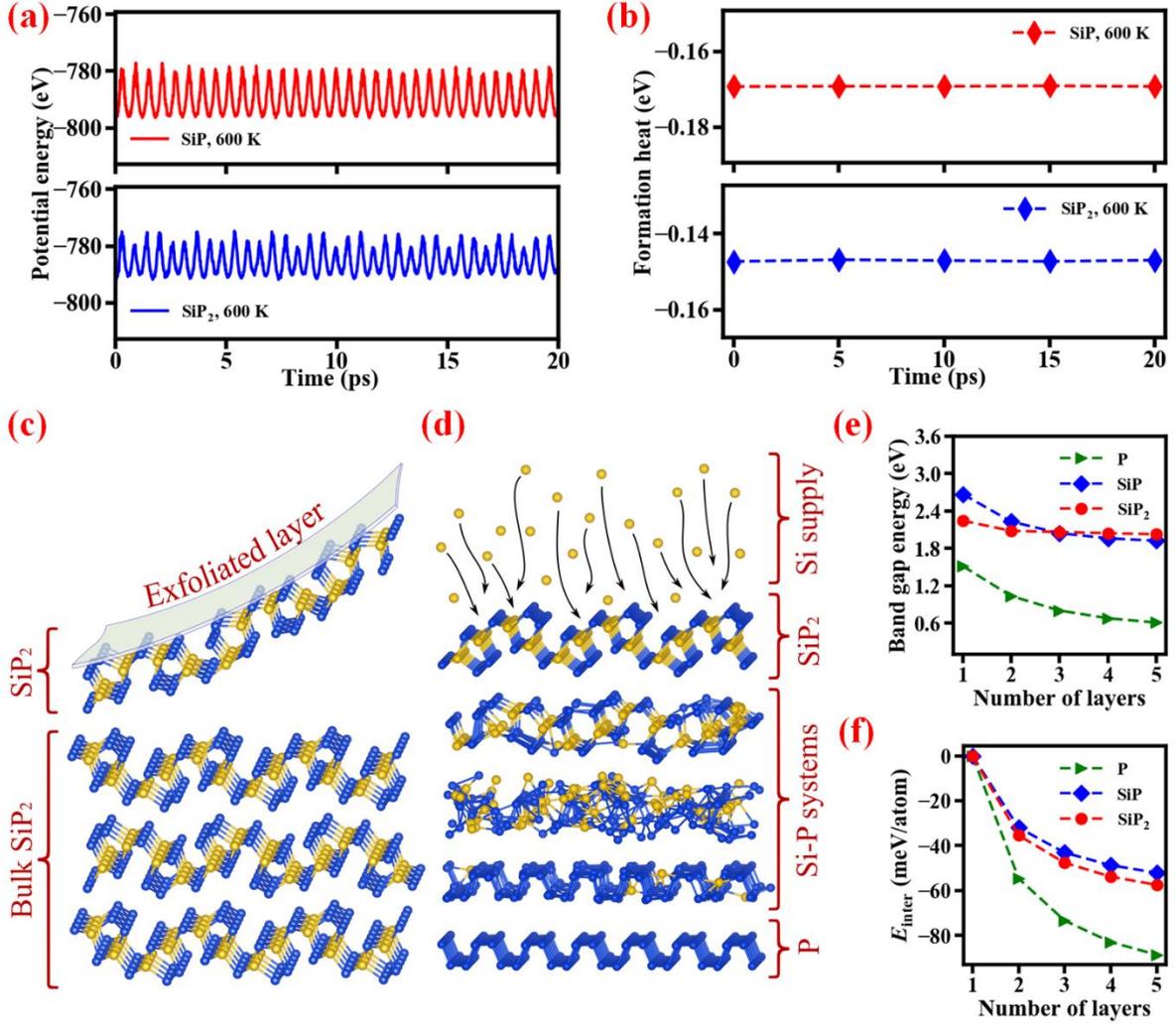

**Figure 5** (a) Evolution of PBE potential energies for a single-layer 2d-SiP and 2d-SiP$_2$ during BOMD simulations at temperature of 600 K. (b) PBE formation heats for structures of a single-layer 2d-SiP and 2d-SiP$_2$ taken from BOMD simulations at different time. Schematic illustration of possible strategies for receiving 2d-SiP$_x$ systems utilizing (c) exfoliation and (d) siliconization. (e) HSE band gap energies and (f) interlayer interaction energies ($E_{inter}$) for 2d-P, 2d-SiP(*Cmc*2$_1$), and 2d-SiP$_2$(*Pnma*) computed for the optB88-vdW relaxed structures.